\documentclass[12pt]{iopart}
\usepackage{graphicx}

\begin{document}

\title{Working with simple machines}

\author{John W. Norbury}

\address{Department of Physics and Astronomy, University of Southern Mississippi, Hattiesburg, Mississippi, 39402, USA}
\ead{john.norbury@usm.edu}
\begin{abstract}
A set of examples is provided that illustrate the use of work as applied to simple machines. The ramp, pulley, lever and hydraulic press are common experiences in the life of a student and their theoretical analysis therefore makes the abstract concept of work more real. The mechanical advantage of each of these systems is also discussed so that students can evaluate their usefulness as machines.

\end{abstract}

\maketitle

\section{Introduction}

The Work-Energy theorem (discussed in detail below)   is one of the most important 
ideas in classical  mechanics, and is often discussed in high school physics courses and
university level freshman physics classes. Nevertheless, the reason for defining work as
force times distance often remains obscure to the student. And the idea that work is
ÔconservedÕ (in the absence of dissipative forces), in that a smaller force implies that the distance must be larger in order
to obtain the same work (and thus impart the same energy to an object), is often lost on
students. On the other hand, simple machines are often studied in elementary school
science classes. These students get to experience how machines can amplify forces. For
instance with a simple lever, the students can lift weights that would otherwise be
impossible. Usually what is emphasized in the study of simple machines is the idea of
mechanical advantage, or force amplification. This is fine and students learn from it.
However rarely is the idea of work emphasized in the study of simple machines.
The idea of the present article is to emphasize the constancy of work in
 the use of simple machines. The aim is twofold. Firstly, it is hoped that after going
through the exercises below, high school and university level physics students will have a
much clearer understanding as to why work is defined as force times distance. Simple
machines will be used to illustrate the definition of work. Secondly, it is hoped that
in the study of simple machines in elementary school, more attention will be paid to the
idea of work. Work will be used to illustrate the use of simple machines.

The outline of the article is as follows. Four different types of simple 
machines (ramp, pulley, lever, hydraulic press) will be studied. In each case it will be
shown explicitly that if less force is applied then a corresponding greater moving
distance is involved such that the work (or effort) always remains the same, and
correspondingly therefore the energy imparted is the same. The theme will be {\em less
force, more distance, same work}. (Or if you like one can say more force, less distance,
same work.)
 A good set of references
on simple machines is provided \cite{ Carin, Harlan, NAP, Lehrman, Harris, Beiser, Bueche, 
Halpern, Hooper, Lea, Fishbane, Hecht}. One of the best is the book by Lehrman \cite{ Lehrman}.
There are also some good sources for elementary school teachers 
\cite{ Carin, Harlan, NAP} and high school and university teachers
\cite{  Lehrman, Harris, Beiser, Bueche, Halpern, Hooper, Lea, Fishbane,
Hecht}.

\subsection{Work-Energy theorem}

	The force approach to the study of mechanics is  to identify all the
forces, divide by mass to get acceleration and then solve for velocity,
displacement, time, etc.  There is an alternative formulation of mechanics
which does not rely heavily on force, but rather is based on the concepts
of work and energy.  The work-energy formulation of mechanics \cite{serway}  is worthwhile
since sometimes it is easier to use  and involves only scalar
quantities.  Also it leads to a better physical understanding of mechanics.
 However the key reason for introducing work-energy is because {\em energy
is conserved}.  This great discovery simplified a great deal of physics.
The basic concept of {\em work} is that   it is {\em force times distance}.
You do work on an object by applying a force over a certain distance.  When
you lift an object you apply a lifting force over the height that you lift
the object.
	{\em Machines}  are  devices that allow   us to do work more efficiently.  For
example, a {\em ramp} is what is called a {\em simple machine}.  If you
load objects into a truck, then a large ramp (large distance) allows you to
apply less force to achieve the same work.
	Actually the proper physical definition of work is more complicated 
than simply force times distance. 
  The
proper definition is  $W\equiv\int_{r_i}^{r_f} {\bf F}\cdot d{\bf  r}$. 
Consider  the 1-dimensional case.
{\em If} the force $F_x$ is {\em constant} then it can be taken outside the
integral to give
\begin{eqnarray}
W &\equiv & F_x\int_{x_i}^{x_f} dx = F_x\Delta x = 
 {\mbox{force $\times$ distance}}
\end{eqnarray}
 where $\Delta x \equiv x_f -x_i$. 
Using  ${\bf  F} = m{\bf  a}$,  and in the 1-dimensional case 
\begin{eqnarray}
W = m\int_{x_i}^{x_f}a\, dx = m\int_{x_i}^{x_f} v \frac{dv}{dx} dx = 
 \frac{1}{2}
mv_f^2-\frac{1}{2} mv_i^2 \equiv \Delta K
\end{eqnarray}
where  {\em
Kinetic Energy} is defined as 
$ K\equiv \frac{1}{2} mv^2$. 
Thus the total work is always equal to the {\em change} in kinetic energy.
Now  recognize that there are two types of
forces called {\em conservative} and {\em non-conservative}.  
To put it briefly, conservative forces ``bounce back'' and
non-conservative forces don't.  Gravity is a conservative force.  If you
lift an object against gravity and let it go then the object falls back to
where it  began.  Spring forces are conservative.  If you pull a spring and
then let it go, it bounces back to where it was.  However friction is
non-conservative.  If you slide an object along the table against friction
and let go, then the object just stays there.
	With {\em conservative} forces we always associate a {\em potential energy}.
	Thus any force ${\bf  F}$ can be broken up into the conservative piece
${\bf  F}_C$ and the non-conservative piece ${\bf  F}_{NC}$, as in
\begin{eqnarray}
W &\equiv& \int_{r_i}^{r_f} {\bf  F}\cdot d{\bf  r}
= \int_{r_i}^{r_f} {\bf  F}_C\cdot d{\bf  r} + \int_{r_i}^{r_f} {\bf 
F}_{NC}\cdot d{\bf  r}
\equiv  W_C + W_{NC}
\end{eqnarray}
and each piece corresponds therefore to conservative work $W_C$ and
non-conservative work $W_{NC}$.   {\em Define} the conservative piece
as the {\em negative} of the change in a new quantity called potential
energy $U$ and  $W_C \equiv -\Delta U$,
giving the {\em Work-Energy theorem},
\begin{eqnarray}
\Delta U+\Delta K = W_{NC}
\end{eqnarray}
If the non-conservative work is zero then we have conservation of total mechanical energy $\Delta E=0$, with $E \equiv U+K$.

\section{Ramp}  The ramp (or inclined plane) is shown in Figure  1, where $h$ is the
 height of the ramp and $s$ is the distance along the ramp. A weight exerts a vertical
force $mg$. For an elementary science class two ramps can be constructed in which the
distance $s$ of one ramp is double the distance of the other. However both ramps should
have the same vertical distance $h$. An example might be loading furniture into a truck
and choosing a ramp to make the job easier. Students can be asked to push the weight up
both ramps. (It is essential  to  put the weight on wheels so that friction is
minimized.) Thus the students will be pushing the weight through the distance $s$. The
force that the students will have to overcome is then $mg sin\theta$. See Figure  1.

\begin{center}
\includegraphics[width=3.5in]{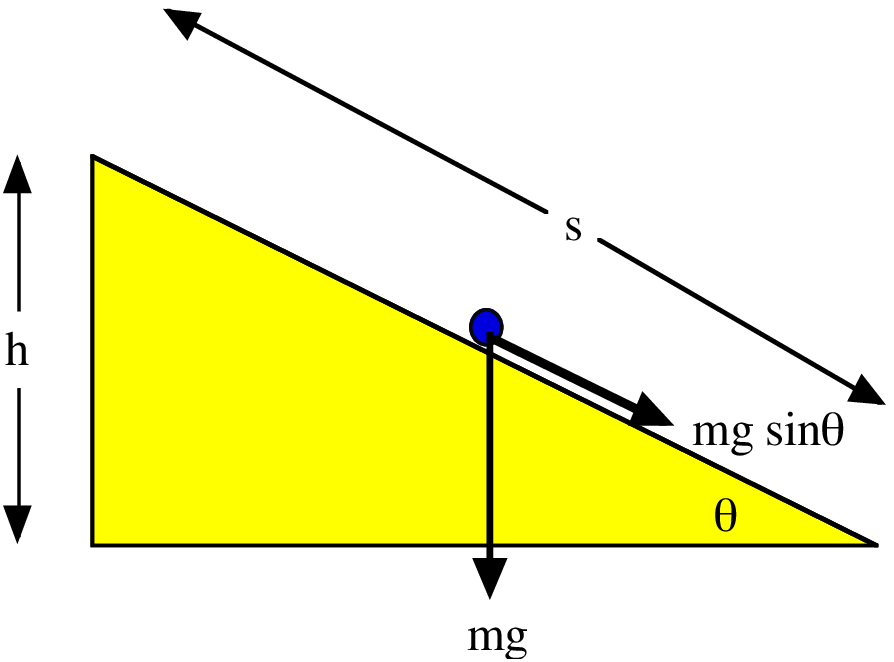}

{\bf Figure  1.}  Ramp.
\end{center}

The essential idea for elementary school students to understand is
 the following. For a long ramp students will notice that the weight is easier to push.
Nevertheless they will be pushing over a correspondingly greater distance and at the end
of pushing the weight up both ramps, they will have expended the same effort. That is
their work is the same. The final potential energy ($mgh$) of the weight is the same.
{\em Less force, more distance, same work.} 

High school and university students can also work this result out
 mathematically as follows. The distance up the ramp is
\begin{equation}
s = \frac{h}{sin \theta}  
\end{equation}
which is large for small angles. The pushing force (see Figure  1) is
\begin{equation}
F=mg sin\theta
\end{equation}
which is small for small angles. Thus ramps are easy to use when the 
angle is small. The product of force and distance is
\begin{equation}
Fs=mgsin\theta \frac{h}{sin \theta} = mgh.
\end{equation}
Thus for both ramps the product $Fs$ is the same. This clearly shows that you can't
 get something for nothing. {\em Less force, more distance, same work.} Students should now have a clear
grasp as to why the work (or effort) is defined as force times distance. It makes sense with the
intuitive idea experienced in actually pushing the weight up the two different ramps. The effort
expended in both cases was the same and this is embodied in the definition of work. (Note that of course
work is really defined as the integral of the scalar product of force and displacement. This
complication is not discussed in this article as the main idea is to develop intuition about the concept
of work.) Most books which discuss simple machines emphasize mechanical advantage rather than work. For
completeness, the mechanical advantage (MA) is defined as the load force divided by the input force or
\begin{equation}
{\rm MA} \equiv {\rm \frac{Load  \;  force}{Input  \; force} }  = \frac{mg}{mgsin \theta } = 
\frac{1}{sin \theta }=\frac{s}{h}.
\end{equation}
Thus a long ramp (large $s$) gives a large mechanical advantage.\\\\

\begin{center}
\includegraphics[width=4in]{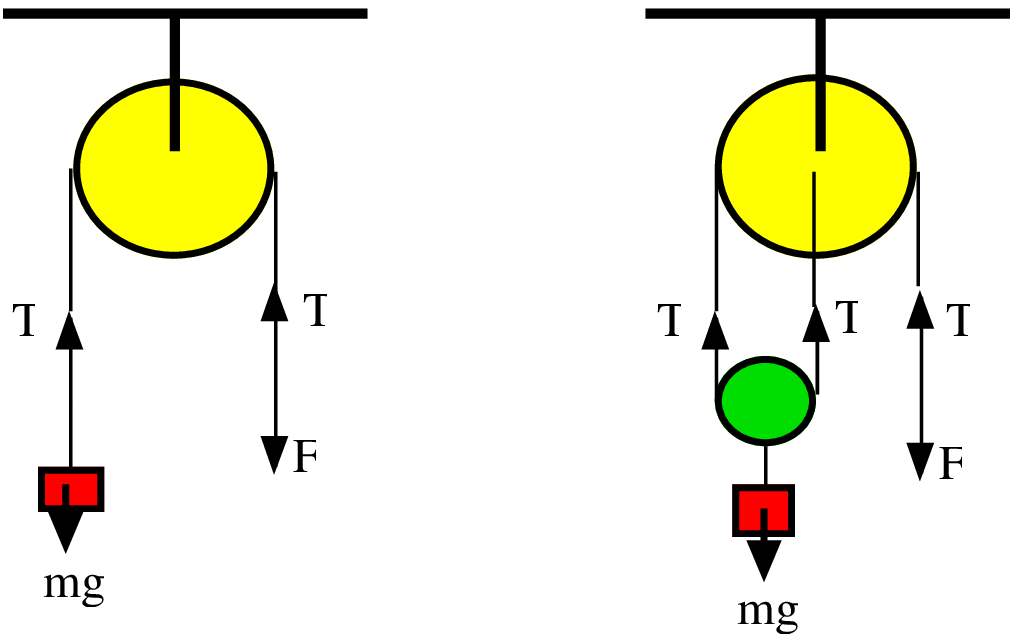}

{\bf Figure  2.}  Pulley Systems.
\end{center}

\section{Pulley}  Next consider some pulley systems as shown in Figure  2. For simplicity 
only consider a single and double pulley. These will be sufficient to illustrate the main idea. Imagine
that the weight is pulled with a force $F$ at constant speed so that the acceleration is zero. Applying
$F=ma$ to the weight gives $T-mg =0$ where $T$ is the tension. Thus the pulling force $F$ equals $T$ and
\begin{equation}
F=mg.
\end{equation}
The rope in the pulley is pulled through a distance $s$. If the weight is 
to be raised by a height $h$ then obviously  
\begin{equation}
s=h.
\end{equation}
Now consider the two pulley system in Figure  2. Again applying $F=ma$ to the 
weight gives $2T-mg=0$ or $T=\frac{mg}{2}$ giving
\begin{equation}
F=\frac{mg}{2}.
\end{equation}
Now by looking at  Figure  2 it can be seen that if the rope is pulled through a
 distance of $s$ then the weight will only be raised by $h=\frac{s}{2}$. Thus now
\begin{equation}
s=2h.
\end{equation}
Thus for the double pulley, it's twice as easy to lift the weight, but one has to
 pull  double the distance to raise the weight the same height. By actually using these pulley
systems the students can experience this for themselves.

The product of force and distance however remains the same.
\begin{equation}
Fs=\frac{mg}{2} 2h = mgh
\end{equation}
{\em Less force, more distance, same work.} Elementary school students can simply 
experience this for themselves by using the two pulley systems. High school and university students can do
the calculations to show explicitly that the product $Fs$ remains the same and thus is good for a
definition of work.

The mechanical advantage is
\begin{eqnarray}
{\rm MA} \equiv {\rm \frac{Load  \;  force}{Input  \; force} }  = \frac{mg}{mg/2 } = 2
=\frac{s}{h}  \\\nonumber
 \\\nonumber
\end{eqnarray}

\begin{center}
\includegraphics[width=4in]{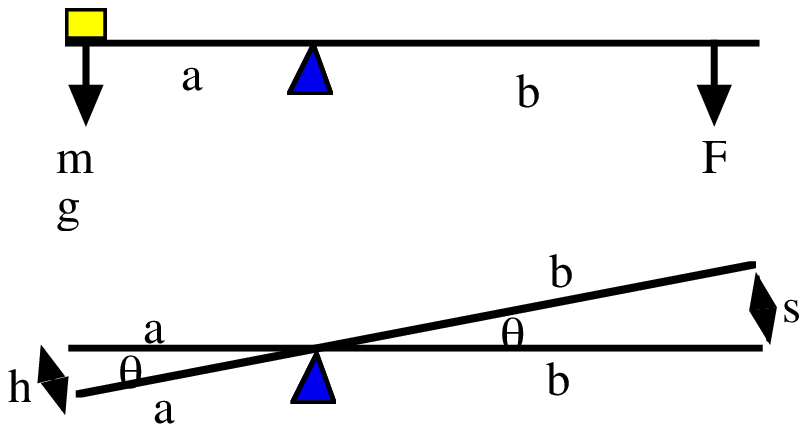}

{\bf Figure  3.}  Lever.
\end{center}

\section{Lever}   The lever is another famous simple machine and is shown in Figure  3. In the top part
of  the figure, a lever is shown in the horizontal position and one imagines that a person is pushing down
with the force $F$ to hold a weight $mg$ in balance. This will be the case no matter what is the
orientation of the lever.  We will ignore the weight of the lever and imagine that we have a lever made of a very light material that does not bend.
For static equilibrium the torques produced by both forces must be the same,
$\tau_1 =
\tau_2$ giving $mga=Fb$ or
\begin{equation}
F=mg\frac{a}{b}
\end{equation}
Thus only a small pushing force $F$ is needed if the lever arm $b$ is large.
 This is something easily demonstrated to students. However if one is to use the lever to lift the
weight, then the longer the lever arm, the larger the distance $s$ one will have to lift through. This
is seen in the bottom part of Figure  3. From the figure $\theta = \frac{h}{a} =\frac{s}{b}$ where $h$ is
the distance the object is to be lifted. The distance that the lever arm will have to be moved through
is 
\begin{equation}
s=h\frac{b}{a}
\end{equation}
showing that if the lever arm $b$ is large (small force) the distance $s$ must be
 large. All students can experience this using the lever. The point is though that the product of force
and distance will always be the same,
\begin{equation}
Fs=mg\frac{a}{b} h \frac{b}{a} = mgh.
\end{equation}
Once again, elementary school students can use the lever , change the length of
 the lever arm and notice that even though it's easier to lift, they have to lift through a larger
distance $s$ and their work, or effort, remains the same. High school and university students can
demonstrate mathematically that the work remains the same. Again we have the principle of {\em less
force, more distance, same work}.

The mechanical advantage is
\begin{equation}
{\rm MA} \equiv {\rm \frac{Load  \;  force}{Input  \; force} } = \frac{mg}{F } = \frac{b}{ a} =
\frac{s}{h}.
\end{equation}

\section{Hydraulic Press} Finally the hydraulic press can also be used to demonstrate these ideas.
 The press is shown in Figure  4. The pressure $P$ is defined as
\begin{equation}
Pressure \equiv \frac{Force}{Area}
\end{equation}
but the pressure throughout the fluid is the same. Thus
\begin{equation}
P = \frac{F}{A_1}= \frac{mg}{A_2}
\end{equation}
giving the applied force as
\begin{equation}
F = \frac{A_1}{A_2}mg
\end{equation}
which shows that for a large area $A_2$ then only a small force $F$ need be 
applied. However from Figure  4  it can be seen that if  $A_2$ is large then the weight $mg$ will only be
lifted a small distance $h$. This can be seen mathematically. The volume change $\Delta V$ will be the
same,
\begin{equation}
\Delta V = A_1 s = A_2 h
\end{equation}
giving
\begin{equation}
s=\frac{ A_2}{ A_1}h
\end{equation}
showing that if $A_2$ is large (small force needed) then the distance $s$ over which 
the force must be applied will have to be large. Students can experience this by using two different
hydraulic presses with different areas $A_2$. The work is always the same. Mathematically this is seen
from 
\begin{equation}
Fs=\frac{ A_1}{ A_2}mg\frac{ A_2}{ A_1}h=mgh
\end{equation}
For the hydraulic press  {\em less force, more distance, same work}. The mechanical advantage is
\begin{equation}
{\rm MA} \equiv {\rm \frac{Load  \;  force}{Input  \; force} }  = \frac{mg}{F } = \frac{ A_2}{
A_1} =
\frac{s}{h}\\
\end{equation}

\vspace*{.5cm}
\begin{center}
\includegraphics[width=4in]{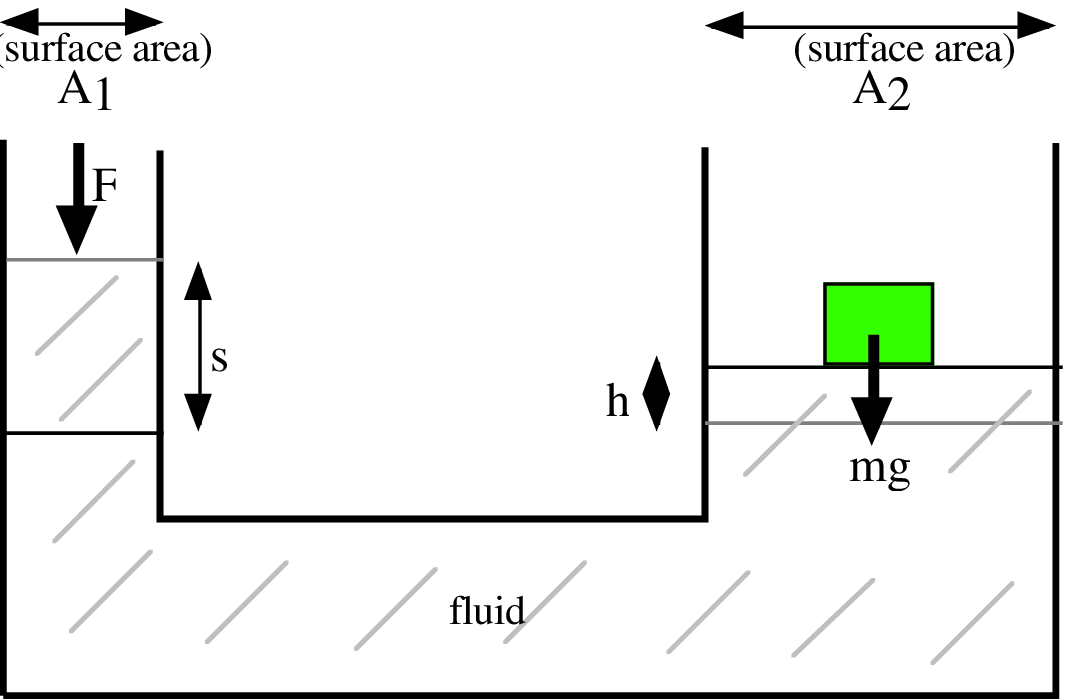}

{\bf Figure  4.}  Hydraulic Press.
\end{center}

In this section we have assumed an idealised hydraulic press. However in reality frictional effects can be rather large and this will make the mechanical advantage smaller than the ideal case.

\section{Conclusion}

The simple machines consisting of a ramp, pulley, lever and  hydraulic press are found in everyday life. Their  mathematical analysis makes the concept of work much clearer to students encountering the subject for the first time. 
These machines can be easily constructed and used as demonstrations in the classroom, which makes them doubly effective for learning. When students interact with these demonstrations they can easily experience the concept of mechanical advantage which has been presented in this paper. Both demonstrations and the analysis of this paper, make simple machines an excellent way for students to learn about work.

\end{document}